# Stacking and Hydrogen Bonding. DNA Cooperativity at Melting.

Vladimir F. Morozov[*], Artem V. Badasyan, Arsen V. Grigoryan,

Mihran A. Sahakyan, Evgeni Sh. Mamasakhlisov.

[1] *Department of Molecular Physics, Yerevan State University, A.Manougian Str.1, 375025,*

*Yerevan, Armenia.*

## Abstract

*By taking into account base-base stacking interactions we improve the Generalized Model of Polypeptide Chain (GMPC). Based on a one-dimensional Potts-like model with many-particle interactions, the GMPC describes the helix-coil transition in both polypeptides and polynucleotides. In the framework of the GMPC we show that correctly introduced nearest-neighbor stacking interactions against the background of hydrogen bonding lead to increased stability (melting temperature) and, unexpectedly, to decreased cooperativity (maximal correlation length). The increase in stability is explained as due to an additional stabilizing interaction (stacking) and the surprising decrease in cooperativity is seen as a result of mixing of contributions of hydrogen bonding and stacking.*

1. Introduction

Since the 1960's the helix - coil transition in biopolymers has been a topic of intensive investigations [1-7] and is still vigorously discussed [8-15]. Traditionally the theoretical models for the transition assume that each base pair can be in either the helical or the coil state. This assumption makes it convenient to use the Ising model [16-19] or to calculate the free energy directly as though the system were a dilute one-dimensional solution of helix and coil junctions [6]. While different in details, most traditional theories use the mean-field approximation. In other words, the Hamiltonians of these models use parameters that are averages over all conformations of the molecule (e.g. the cooperativity parameter in Zimm - Bragg theory [16-19] or the junction free energy [6]). The microscopic theory of Lifson et al. for polypeptides [20-22] and the more recent Peyrard and Bishop theory for DNA [11-13] do not make the mean field assumption.

---
[*] Corresponding author e-mail: morozov@ysu.am

It is widely accepted that the helical structure of DNA is conditioned by the presence of two types of interaction. The first type, known as stacking, restricts the conformational states of nearest-neighbor base pairs. It is believed that this type of interaction models the hydrophobic attraction between nearest-neighbor base pairs [1, 2, 5]. The stacking is explained in following way: the hydrophobicity of nitrogen bases, flat heterocyclic compounds, causes parallel packing like stack of coins. The second interaction is the hydrogen bonding of complementary base pairs. Hydrogen bonding restricts the conformational states of repeated units on a finite scale, larger than nearest-neighbors'. Not only the stability of DNA conformations but also the correlation (which results in cooperativity of helix - coil transition [3]) is conditioned by these interactions. Consider how these interactions are modeled in two typical theories of the transition in polypeptides. In Zimm-Bragg theory stacking is modeled as nearest-neighbor attraction; the cost of junction between helical and coil regions [2, 14, 16-19] causes the cooperativity. In Lifson - Roig theory the restrictions on chain backbone conformations, imposed by hydrogen bond formation are taken into account [15, 20-22] and model the cooperativity. The results of both theories do not differ greatly [15], i.e., each factor separately results in cooperativity. To investigate the simultaneous influence of these two interactions a microscopic theory should be applied [15]. The importance of investigation of joint contribution of hydrophobicity and hydrogen bonding was also discussed in [23].

Because the simultaneous influence of stacking and hydrogen bonding on cooperativity has not yet been considered, we investigate this problem within the content of our Generalized Model of Polypeptide Chain (GMPC) and reveal the role of stacking against the background of hydrogen bonding.

2. Basic Model (GMPC).

A microscopic Potts-like one-dimensional model with $\Delta$-particle interactions describing the helix – coil transition in polypeptides was developed in [24, 25]. Then it was shown that the same approach could be applied to DNA if ignore large-scale loop factor [26]. The Hamiltonian of GMPC has the form of the sum over all repeated units:

$$-\beta H = J \sum_{i=1}^{N} \delta_i^{(\Delta)}, \qquad (1)$$

where $\beta = T^{-1}$ is inverse temperature; $N$ is the number of repeated units; $J = U/T$ is the temperature-reduced energy of interchain hydrogen bonding; $\delta_j^{(\Delta)} = \prod_{k=\Delta-1}^{0} \delta(\gamma_{j-k}, 1)$, with Kronecker $\delta(x,1)$; $\gamma_l$ is spin that can take on values from 1 to $Q$ and describes the conformation of $l$-th repeated unit. The case when $\gamma_l$ is equal to 1 denotes the helical state, other $(Q-1)$ cases correspond to coil state. $Q$ is the number of conformations of each repeated unit and thus describes the conformational ability. The Kronecker delta inside the Hamiltonian ensures that energy $J$ emerges only when all $\Delta$ neighboring repeated units are in helical conformation. Thus the restrictions on chain backbone conformations, imposed by hydrogen bond formation, are taken into account [26].

The transfer - matrix, corresponding to the Hamiltonian Eq.(1) looks like:

$$\hat{G}(\Delta) = \begin{pmatrix} V & V & V & \ldots & V & V & V \\ 1 & 0 & 0 & \ldots & 0 & 0 & 0 \\ \ldots & \ldots & \ldots & \ldots & \ldots & \ldots & \ldots \\ 0 & 0 & 0 & \ldots & 1 & 0 & 0 \\ 0 & 0 & 0 & \ldots & 0 & 1 & Q \end{pmatrix}, \qquad (2)$$

where all elements of first row are equal to $V = \exp(J) - 1$; all elements of first lower pseudodiagonal are 1; the $(\Delta, \Delta)$ element is $Q$; all other elements are zero. The secular equation for this matrix is:

$$\lambda^{\Delta-1}[\lambda - (V+1)](\lambda - Q) = V(Q-1) . \tag{3}$$

As previously shown [25], the two-particle correlation function of this model in thermodynamic limit can be written as

$$g_2(r) = \langle \delta_i^{(\Delta)} \delta_{i+r}^{(\Delta)} \rangle - \langle \delta_i^{(\Delta)} \rangle \langle \delta_{i+r}^{(\Delta)} \rangle \sim \exp\left[-\frac{r}{\xi}\right], \tag{4}$$

where $r$ is the distance (in repeated units), and

$$\xi = [\ln \lambda_1 / \lambda_2]^{-1} \tag{5}$$

is the correlation length; $\lambda_1$ is the largest eigenvalue, $\lambda_2$ is the second largest. Near the transition point, estimated from $W = Q$ condition as $T_m = U / \ln Q$, the correlation length $\xi$ passes through the maximum, which can be estimated as

$$\xi_{max} \sim Q^{\frac{\Delta-1}{2}} \tag{6}$$

[24-26]. The parameter $\sigma$ of Zimm-Bragg theory corresponds with $\xi_{max}$ [26] as

$$\sigma = \xi_{max}^{-2} . \tag{7}$$

The following set of parameters was estimated for DNA in Ref.[26]: $Q \propto 3 \div 5$; $\Delta \propto 10 \div 15$. One can see that for this set the cooperativity parameter $\sigma \sim 10^{-5}$ to $10^{-7}$. So, the high cooperativity of homogeneous DNA was explained as determined by large value of $\Delta$.

3. Model With Stacking

Our base model considers cooperativity through the hydrogen bonding, while in some approaches the cooperativity is determined through stacking interactions [1, 2, 5]. Taking stacking into account, we can write the Hamiltonian with stacking by analogy with Hamiltonian Eq.(1) as

$$-\beta H = J\sum_{i=1}^{N}\delta_i^{(\Delta)} + I\sum_{i=1}^{N}\delta_i^{(2)}. \tag{8}$$

The first term on the rhs is the same Hamiltonian Eq.(1), describing $\Delta$- range interactions. The second term describes nearest-neighbor-range interactions (stacking fixes in helical conformation two nearest-neighbor repeated units). Here $I = E/T$ is the reduced energy of stacking interactions. The Kronecker $\delta_i^{(2)}$ ensures that the reduced energy $I$ is emerged when two nearest neighboring repeated units are in the same, helical conformation. The transfer - matrix for the model with the Hamiltonian Eq.(8) looks like

$$\hat{G}(\Delta) = \begin{pmatrix} VR & VR & VR & ... & VR & VR \\ R & 0 & 0 & ... & 0 & 0 \\ 0 & R & 0 & ... & 0 & 0 \\ ... & ... & ... & ... & ... & ... \\ 0 & 0 & 0 & ... & R-1 & R-1 \\ 0 & 0 & 0 & ... & 1 & Q \end{pmatrix}, \tag{9}$$

where $R = \exp[I]$; $VR = W - R$; $W = \exp[J+I]$. It is obvious, that at $R=1$ Eq.(9) passes into Eq.(2). The structure of transfer - matrix (9) is rather similar to (2). The principle difference is that in the right lower corner of ($\Delta \times \Delta$) matrix there is some ($2 \times 2$) matrix, which corresponds to the base model with $\Delta=2$. The secular equation for the transfer - matrix looks like:

$$\lambda^{\Delta-2}(\lambda-W)\left[\lambda^2 - (R+Q-1)\lambda + (R-1)(Q-1)\right] = R^{\Delta-1}V(Q-1). \tag{10}$$

In principle, as far as we have characteristic equation we can obtain eigenvalues (at least numerically) and then calculate the main quantity of any helix-coil transition theory, namely, the helicity degree as

$$\theta = \frac{\partial \ln \lambda_1}{\partial J}. \tag{11}$$

By analogy the average fraction of stacked repeated units can be obtained as well:

$$\varepsilon = \frac{\partial \ln \lambda_1}{\partial I}. \qquad (12)$$

This is the matter of further investigation. But in this given article we are not constructing the theory of helix-coil transition with both stacking and hydrogen bonding. The problem of interest is how stacking (on the background of hydrogen bonding) affects the stability and cooperativity of DNA. The most convenient way to solve this problem is to calculate of correlation length using eigenvalues from Eq.(10). That is why in the same way as in the basic model, we introduce a two-particle correlation function as Eq.(4) with correlation length as Eq.(5). The calculation shows that in analogy with the basic model the temperature dependence of correlation function has a maximum. The temperature at this maximum corresponds to the transition point; and the maximal value of correlation length characterizes the cooperativity of transition. Introducing $\alpha = \frac{E}{U}$, the energetic contribution of stacking against the background of hydrogen bonding, we calculate the dependence of the correlation length on the temperature parameter $W = \exp\left[\frac{U}{T}(1+\alpha)\right]$ for $\alpha$ ranging from 0 to 2. The results presented in Fig. 1 are in dimensionless units $\frac{\xi}{\xi_0}, W$; where $\xi_0$ is the maximal correlation length with $\alpha = 0$ the case, i.e. of the basic model without stacking. Fig. 2 represents the behavior of dimensionless melting temperature $\frac{T_m}{T_{0m}}$ ($T_{0m}$ is the melting temperature of basic model) on $\alpha$. The dependence of the dimensionless maximal correlation length on $\alpha$ is shown in Fig.3.

4. Discussion.

In Fig.1 one can see, that raising $\alpha$ shifts curves to the right, lowers the maxima and makes the curves wider. This shift means that the maximum of correlation length at a given $\alpha$ corresponds to values of $W > Q$. The calculations show that the shift in the region of $\alpha \in [0,2]$ is practically linear in $\alpha$. Therefore the condition for a transition point may be written

$$\frac{\exp\left[\frac{U}{T}(1+\alpha)\right]}{Q} = 1 + c\alpha \qquad (13)$$

with constant $c > 0$.

In Fig. 2 one can see that melting temperature increases linearly in $\alpha$. The increase in stability is the result of stacking energy added to the Hamiltonian. The slope of this curve is less than unity. It happens due to the shifts of maximums in Fig. 1. To see the source of this slope and linearity, expand Eq.(13) around $T_{0m}$ ($\alpha$ close to zero)

$$\frac{T_m}{T_{0m}} = 1 + \alpha\left(1 - \frac{c}{\ln Q}\right). \qquad (14)$$

Fig. 2 shows, that this linearity holds up to $\alpha = 2$. The linear behavior of the transition temperature explains why mean field theories [17-19], in which the helix stabilization energy is additive sum of contributions of different mechanisms of stabilization, describe the experimental data on melting temperature so well over vast range of energies.

Now consider correlations. In Fig's 1 and 3 one can see that the maximal correlation length decreases with stacking energy, i.e. $\alpha$, and the increased stacking energy relative to the hydrogen bonding energy results in decreased range of correlation. This result was unexpected for us, because in the Hamiltonian Eq.(8) the term with nearest-neighbor correlation was introduced in addition to the $\Delta$ correlated one. It seemed that introducing an additional helix-stabilizing interaction would result in an increased maximal correlation length as well as it resulted in increased melting temperature, but it was not so.

At $\alpha = 0$ we deal with the pure basic model with range of correlation $\Delta$; when $\alpha \to \infty$ we deal with range of correlation, equal to two, typical of stacking. As one can see from Eq.(6), the maximal correlation length for the basic model at $\alpha = 0$ is much larger than at $\alpha \to \infty$. It is obvious, that at intermediate $\alpha$'s the maximal correlation length will take on intermediate values, i.e. the situation, presented in Fig. 3. So we have some mixing of maximal correlation length between the $\Delta$ - correlated and the nearest-neighbor correlated cases. This explains why increasing stacking energy vs hydrogen bonding results in the decreased correlation scale of the system.

It is widely accepted that the main cause of cooperativity is stacking. However it should be kept in mind, that the comparative analysis of two cooperativity factors, stacking and hydrogen bonding, can not be performed [15] with the frequently used mean field approximation. So we must compare our results with those of another microscopic theory of DNA melting which takes into account the difference in the mechanisms of hydrogen bonding and stacking.

Consider another microscopical approach, namely, the Peyrard-Bishop approach [11-13]. This model [11-13] recognizes both stacking and hydrogen bonding. It uses the analogy between the statistical treatment of macromolecules and the Schroedinger equation [2, 27] and treats the problem as a particle in Morse potential which reflects hydrogen bonding. Harmonic coupling reflecting stacking is assumed between nearest neighbor repeated units. The case of disorder in sequence [10] was studied as well. The increase in melting temperature and the decrease in melting interval follows from Fig.1 of Ref.[11]. This Figure compares the calculated melting curves for different contributions of stacking. The authors discuss only melting temperature and compare it with Zimm-Bragg mean-field theory. In the same picture the increased melting interval at increased stacking is shown, but there are no comments concerning this point.

The Peyrard-Bishop results coincide with ours. In both models the difference of mechanisms of stacking and hydrogen bonding interactions in DNA is correctly introduced. In the Peyrard-Bishop approach this difference consists in the difference of interaction potentials. In our model this difference consists in the different correlation scales, as directly follows from DNA structure.

As shown earlier [24, 25] the action of water and other solvents may be included in our basic model by redefinition of model parameter $W$ or $Q$. The choice of redefined parameter depends on the mechanism of solvent interaction. This account makes our qualitative analysis applicable to DNA melting experiments.

As is known from the literature, the contributions of stacking and hydrogen bonding into the energy of helical state are of the same order in water [28]. Therefore in real DNA we deal with the case $\alpha \sim 1$, which corresponds to significant lowering of the maximal correlation length compared to the case $\alpha = 0$ (pure basic model). This lowering is by twenty times (Fig.3). So we suggest that the increased role of stacking (or the decreased role of hydrogen bonding) will result in decreased cooperativity.

Acknowledgments
Authors thank Dr Adrian Parsegian for useful discussions and for help improving the style and language of this paper. This work was supported in part by ISTC Grants No. A-301.2, No. A-092.2.

Figure Caption

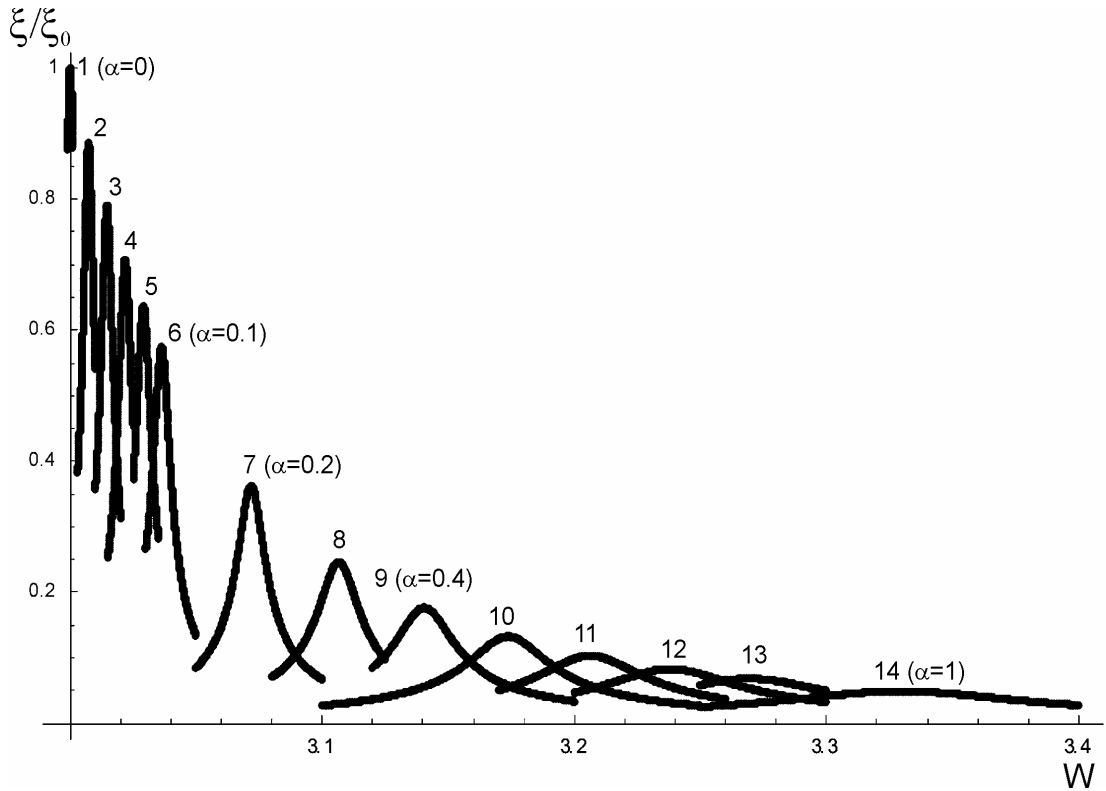

Figure 1. The dependencies of reduced correlation lengths $\xi/\xi_0$ on temperature parameter $W$ for fixed values of $\alpha = \dfrac{E}{U}$. The curves are enumerated corresponding to the following values of $\alpha$ : 1 – 0; 2 – .002; 3 – .004; 4 – .006; 5 – .008; 6 – .1; 7 – .2; 8 – .3; 9 – .4; 10 – .5; 11 – .6; 12 – .7; 13 – .8; 14 – 1.

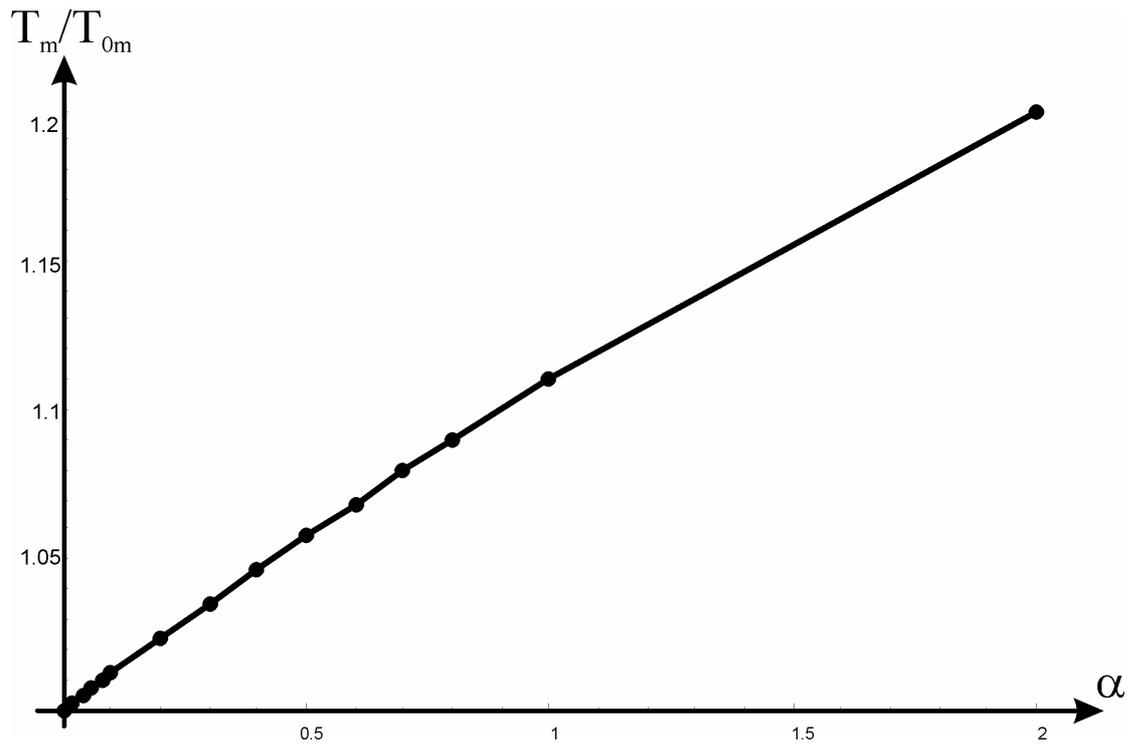

Figure 2. The dependence of reduced maximal correlation length $\xi_{max}/\xi_{0\,max}$ on $\alpha$.

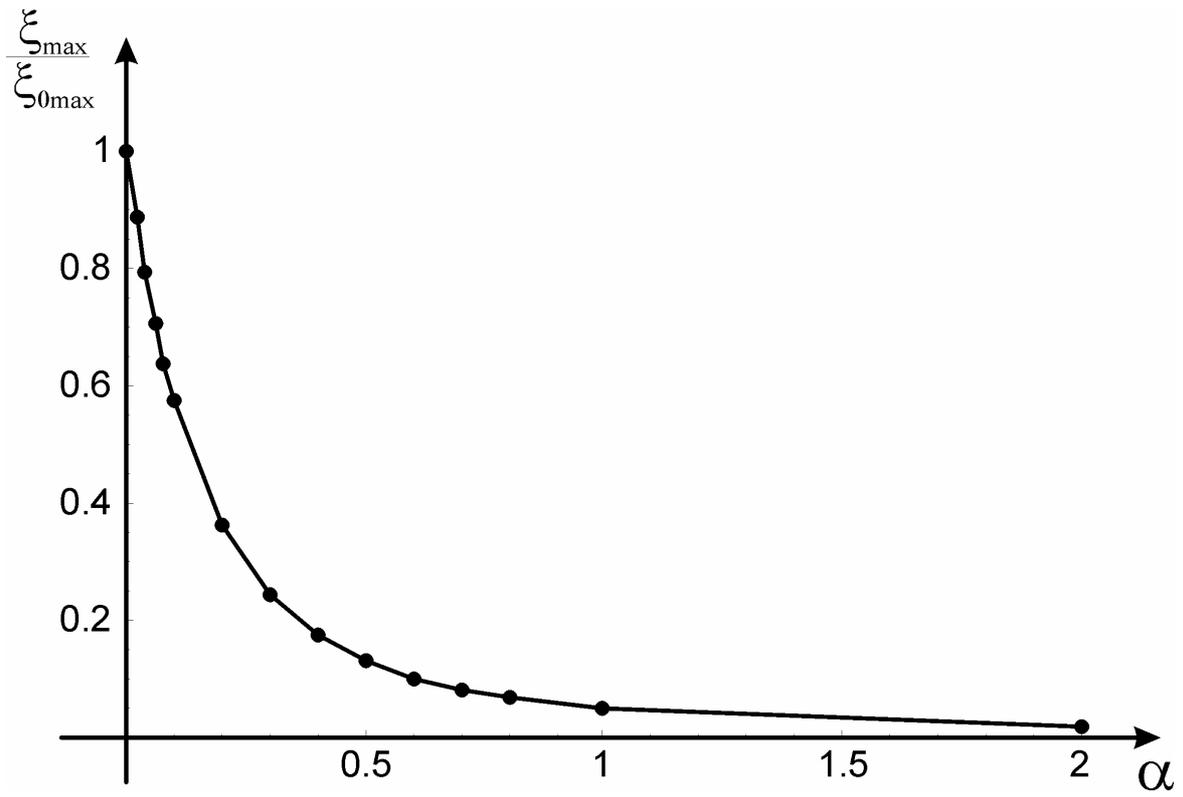

Figure 3. The dependence of reduced transition temperature $T_m/T_{0m}$ on $\alpha$.